# Privacy and Security Improvement in UAV Network using Blockchain

Hardik Sachdeva[a],   *Shivam Gupta[b],   Anushka Misra[a],   Khushbu Chauhan[c],   Mayank Dave[a]

[a]National Institute of Technology (NIT), Kurukshetra , India.
[b]Indian Institute of Technology (IIT), Ropar, India.
[c]Indian Institute of Information Technology (IIIT), Sonepat (Mentor & Campus National Institute of Technology Kurukshetra), Haryana, India.

*Correspondence addressed to: shivam.20csz0004@iitrpr.ac.in

**Abstract** – Unmanned Aerial Vehicles (UAVs), also known as drones, have exploded in every segment present in today's business industry. They have scope in reinventing old businesses, and they are even developing new opportunities for various brands and franchisors. UAVs are used in the supply chain, maintaining surveillance and serving as mobile hotspots. Although UAVs have potential applications, they bring several societal concerns and challenges that need addressing in public safety, privacy, and cyber security. UAVs are prone to various cyber-attacks and vulnerabilities; they can also be hacked and misused by malicious entities resulting in cyber-crime. The adversaries can exploit these vulnerabilities, leading to data loss, property, and destruction of life. One can partially detect the attacks like false information dissemination, jamming, gray hole, blackhole, and GPS spoofing by monitoring the UAV behavior, but it may not resolve privacy issues. This paper presents secure communication between UAVs using blockchain technology. Our approach involves building smart contracts and making a secure and reliable UAV adhoc network. This network will be resilient to various network attacks and is secure against malicious intrusions.

**Keywords:** Unmanned Aerial Vehicles (UAVs), Blockchain, Data Privacy, Network Security, Smart Contract, Ethereum.

## 1.    Introduction

UAV is an aircraft that can steer without a human pilot onboard the aerial vehicle. UAVs started as a cost-effective alternative to human-crewed military aircraft and are likely to continue in the future with improvement in technology. Like the internet and GPS, UAVs are progressing beyond their defense applications to become helpful business tools in the civilian domain due to the recent advancements in their functioning, network technology, communication, and manufacturing processes [30-34]. They are getting into government and commercial services [49]. It has ended up creating a tremendous market opportunity for the industry [4].

The increasing popularity of UAVs is exposing their limitations too. The programming languages used to develop software for UAVs are not intentionally proposed for the objective and thus are prone to hackers to crack due to bugs in the languages [5]. Also, UAVs are prone to be lost, physically hijacked, or destroyed because of deployment in an open atmosphere [35, 37-38].. With the UAV technology becoming global, various issues arise in UAV networks that need addressing, such as UAV security, management and storage of data, intra-UAV communication, and air data security. UAV ad-hoc network (UAANET), as shown in Figure 1, comprises UAVs and base stations [45-47]. . Base stations are also known as ground control stations (GCS). The terms  UAV network and UAANET are used interchangeably in the paper. The UAVs and GCSs register with a central trusted authority known as the control room (CR) before their deployment. The drones and the GCS communicate over open wireless channels, leading to many security and privacy issues in their environment [1,2]. Thus, it becomes mandatory to ensure that the communication and transmission of data within a UAANET is not interrupted or disrupted by malicious entities.

One can apply a suitable and secure technology such as blockchain to provide a defense mechanism against the increasing number of cyber-attacks in the UAV network [3, 36]. Using this technology, we can communicate within a UAANET more securely [48]. Each node in the network has a copy of all the data as blockchain is a distributed ledger. A blockchain network can be corrupted or destroyed if the hacker attacks or destroys each UAV present in the blockchain network. The hacker can't take down an entire network [39]

. The use of blockchain prevents the entry of a malicious node into the network, and data security is enhanced.

The present work tries to improve the UAV network's privacy and security using blockchain technology. We also develop a novel simulator for UAV networks that a wireless remote controller controls. The main contributions of the paper are summarized below.

***Contributions***:
1. We use encryption and decryption while sending and receiving the data to ensure data privacy. The use of asymmetric cryptography allows only the destination to access the data.
2. To detect an attack, we keep track of all blockchain transactions (data transmitted by each node, timestamp of all transactions, and routing table of each node). One cannot change the transaction information on blockchain due to its immutable nature, thus ensuring data integrity.
3. To establish trust among the participating network nodes, there is an exchange of tokens among them in the blockchain. And for ensuring the authenticity of the route to the source node, the intermediary nodes pay Ethers as a guarantee for successful transmission.

Thus, the work aims to increase network security by detecting and preventing various attacks in the UAV network by using blockchain.

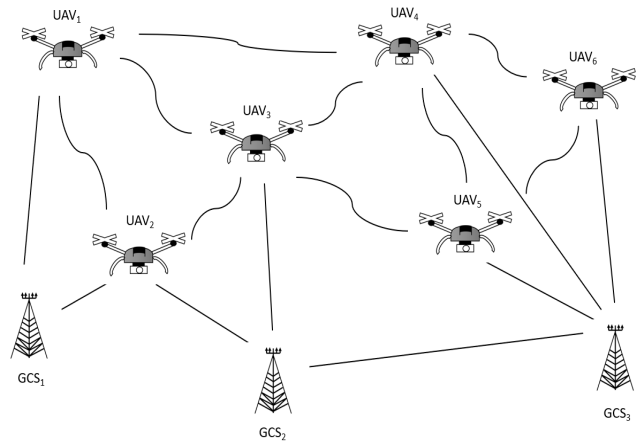

**Figure 1.** A UAANET consisting of UAVs and GCSs communicating over an open wireless channel

## 1.2 Paper Positioning

The remainder of this paper is organized as- Section 2 discusses the related issues and characteristics of blockchain for the proposed design. Section 3 highlights the proposed approach and discusses the concept of node registration in the UAV Network, the flow of data transactions and contract functions in the blockchain, and the simulation of UAANET in Python. Section 5 presents the implementation details and the underlying algorithms of the designed system. Results and observations drawn from the proposed approach simulation are thoroughly in section 6. Finally, follow the paper's conclusion and possible future directions.

## 2. Preliminaries

This section discusses the following preliminaries that are essential to describe and understand the proposed scheme.

### 2.1 Unmanned Aerial Vehicles (UAVs)

The UAV system consists of sensor-payloads, aircraft components, and a GCS. One can control UAVs using control equipment from the ground or onboard electronic equipment [11]. The UAVs collaborate to relay data for transmission from source to destination, control and command the traffic, and remotely sense the UAVs and the GCS [6]. One can improve the security of UAANET by handling the challenges of the CIA triad (Confidentiality, Integrity, and Availability). To overcome such challenges, one can use blockchain technology discussed later. Among specific features of UAV, one important point is related to the number of nodes in a UAANET. In line with previous literature, these are limited to 3-4 as they are considered sufficient for applications under consideration [7,8,9]. The elementary requirements for communication and flying ad hoc networks (FANETs) are explored with coordination, device mobility, and control which also requires certification on the deployment of several UAVs [2,5].

### 2.2 Blockchain Technology

A blockchain is a collaborative, tamper-resistant ledger that maintains transactional records grouped into blocks [16]. A block becomes permanent in the blockchain after transaction verification [17]. Every block connects to its previous block by using a unique identifier. Any change in the data block leads to modifying a unique identifier, and all users get informed about the change. The nodes reject all such tampered blocks. Thus the blockchain network is challenging to alter or destroy, a resilient method of collaborative record keeping [16].

The immutable and distributed property with no centralized authorization enhances its security [16,17]. In this paper, public key infrastructure (PKI) is used in blockchain to encrypt data [19]. To automate dynamic UAV systems, one can use consensus mechanisms and smart contracts like present work. It is motivated by traceability and automated execution of business logic features of blockchain [20]. Asymmetric encryption ensures the authentication of the signature of the corresponding UAV [21]. The use of blockchain technology efficiently decreases the possibility of data change by malicious, illegal parties [22].

### 2.3 Cyber-attacks on UAV Network

The UAANET is susceptible to various threats and cyber-attacks due to its inherent characteristics of UAVs. Adversaries may cause destruction and loss of data by exploiting the radio waves. One can achieve this by adding malicious nodes, controlling and absorbing the network traffic, and disrupting the routing functionality [6]. Other cyber-attacks UAVs are susceptible to are Blackhole attack [12], Wormhole attack [13], Denial of Service (DoS) attack, Sybil attack, and Byzantine attack [6].

## 3. Related Works

The work in developing secure UAVs is in the nascent stages. In recent years, a couple of cyber-attack detection and response system schemes have come up. The authors in [14] describe a lightweight scheme that aims to detect the occurrence of cyber-attacks such as GPS spoofing, false information dissemination, and jamming. A rule-based intrusion detection scheme has been proposed by [13]. The scheme identifies attacks where each node activates as an intrusion detection agent, i.e., UAV detection agent (UDA), in monitoring mode. It helps UDA in hearing all packets in its radio range. To detect false information dissemination attacks, every UAV keeps a check on the physical phenomenon of its neighboring UAVs like injured persons, traffic accidents, forest fires, etc. The main limitation of the works is that a malicious node can still change the sensor's reading and thus inject false information [15]. Also, the UDA compares the observed phenomena with those broadcasted by its other UAV neighbors, that once can be manipulated easily as there is no record keeping.

A prospective solution for trust management is the blockchain network, which has been present and active in various research fields like wireless networks [22] and the IoT [50,51]. The UAV adhoc network's resource limitation is essential for designing a trust management system that benefits from the decentralized blockchain. Several researchers have even applied blockchain in drone applications like Package Tracking System by Walmart [5], Drone Delivery by Dorado platform [5], Drone Package Delivery [5], etc., which are some of the popular projects in the field of blockchain technology and drones.

The UAV adhoc network's resource limitation is essential for designing a trust management system that benefits from the decentralized blockchain.

## 4. Proposed Methodology

To increase security in the UAV network and prevent the various attacks in the network, we build smart contracts that handle the data transmission. As smart contracts are immutable and decentralized thus, making the UAV network secure for data transmission and communication. Our system can prevent blackhole attacks, gray hole attacks, DoS attacks, confidentiality attacks, and integrity attacks. A node could be a UAV or a GCS. In the proposed system, if a malicious node drops the data packet or tries to disseminate the data before forwarding it, we detect the malicious node present in the network and penalize the node. On re-occurrence of malicious activities in the network, the node is removed from the network and can no longer participate in the transactions, thus making the system more resilient to cyber-attacks. The pseudo algorithm for the proposed approach is presented in Algorithm 1. and 2. The details will be discussed in this section.

### 4.1 Registration of a new node in the network

When a node wants to join the UAV adhoc network, it sends its blockchain address to the registration contract. If the node is registered already in the network or if it is not registered but blacklisted from earlier then it cannot register again. A node not blacklisted gets registered in the network and its information gets saved in the contract. Figure 2. illustrates the registration of a new node in the network.

### 4.2 Transactions in the network

To make a transaction that involves the transmission of data between UAVs or GCSs or UAV-GCS, two phases are involved -

**Phase 1**: Fetching the nodes and updating the graph

In Figure 3, $node_1$ acts as the source node in a network. The source node triggers a transaction which informs the other nodes present in the network that it wants to initiate a transaction. Nodes in the network keep polling for transaction requests (Figure 3). In case there is a request, the nodes can participate in the network.

The participating nodes are validated first for their history of malicious activity. If a participating node has a bad history then the node has to pay the penalty to further participate in the transactions else the node will be rejected. The graph is updated based on the coordinates provided by the participating nodes.

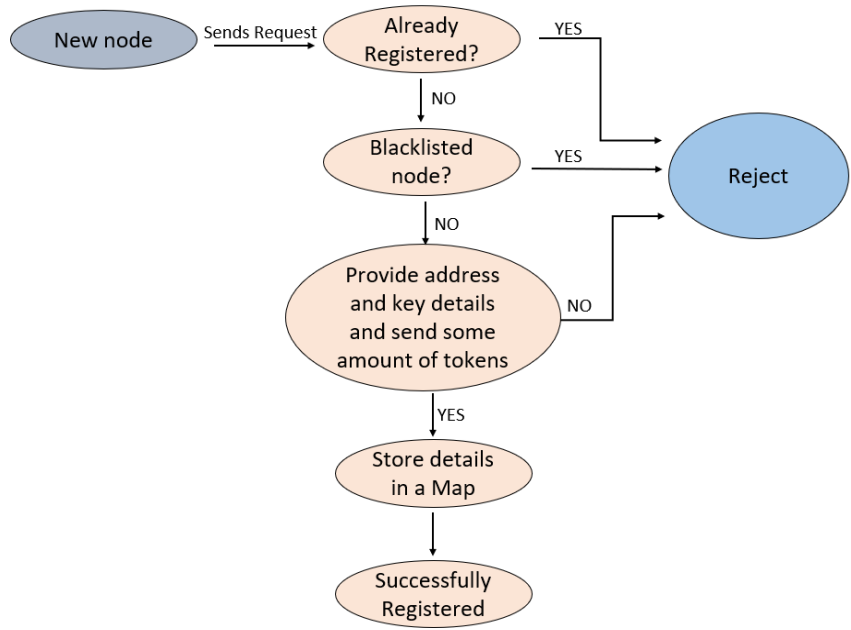

**Figure 2.** Flowchart of the registration of a new node in the network

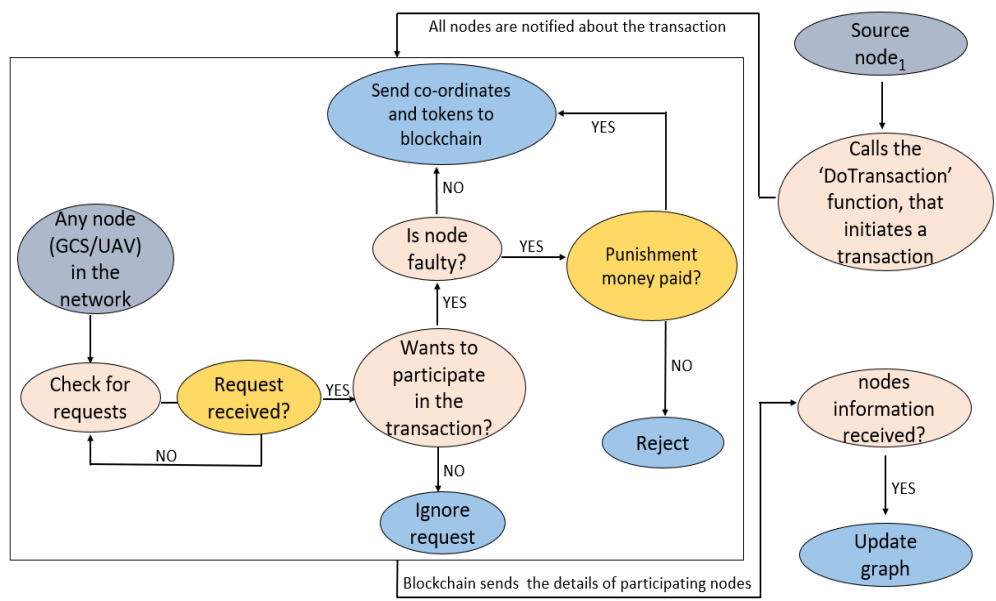

**Figure 3.** Flowchart depicting a transaction between the source node and the destination node (Phase 1)

**Phase 2:** Data sending and detection of the malicious node (if present)

In Figure 4, we have considered $node_1$ as the source. The route to the destination $node_N$ contains the following nodes: $node_2$, $node_3$, ..., $node_{N-1}$.

In the second phase, a path gets generated using the graph formed in Phase 1, and when an optimal path is discovered, the source node sends the data to the next node in the route. Each intermediate node forwards the data packets to the next node. Since any node can become malicious it can even disseminate the data or drop the data packets:

1) Absence of malicious node

In the absence of a malicious node, a transaction is flagged successful by the destination. the tokens submitted by the participating nodes before the transaction as a guarantee get returned. The source node sends appreciation tokens to the participating nodes.

2) Presence of a malicious node

In the case of a malicious node;-a) if the data disseminates, the destination cannot decrypt the message upon receiving the data packets, declaring that the transaction is unsuccessful. The contract then finds the malicious node based on Algorithm 2 and the node is declared faulty and penalized.

b) if a malicious node drops the data packet or the destination stops receiving data packets within an estimated time frame, the destination declares the transaction unsuccessful. The contract then detects the node that did not send data and is declared faulty and penalised. The penalty increases exponentially based on Algorithm 2. A node is blacklisted after a certain number of bad behaviour was shown.

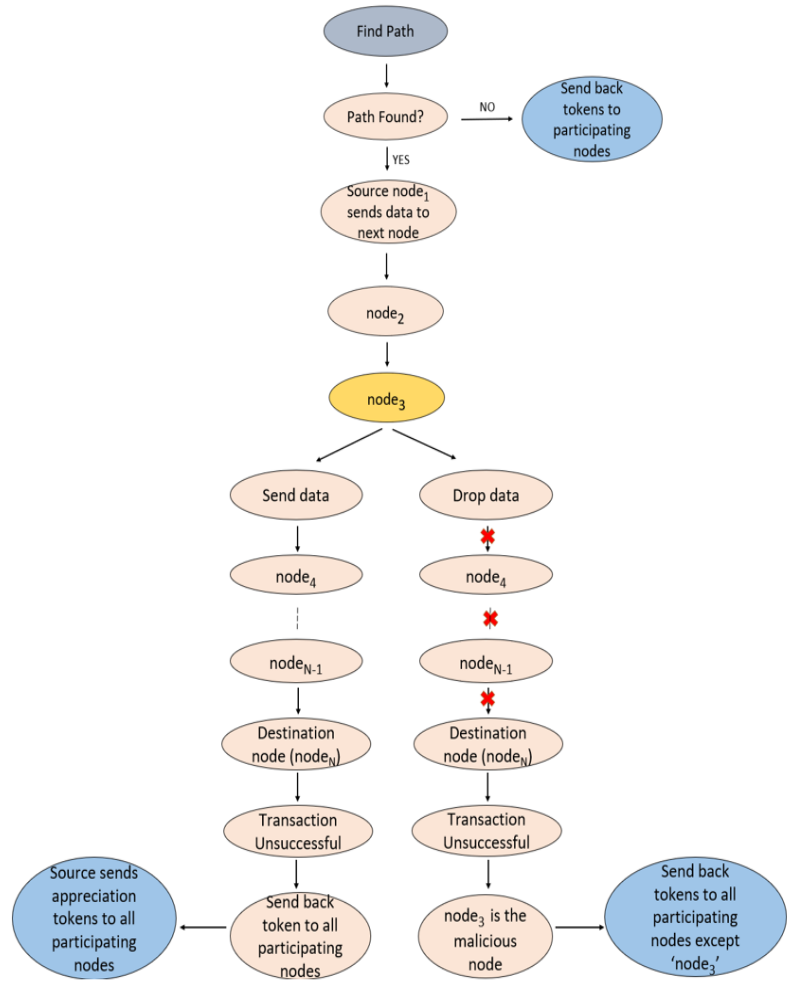

**Figure 4.** Flowchart depicting a transaction with and without the presence of a malicious node in the network

### 4.3 Real-time control of UAVs

The real-time control of UAVs involves plotting the updated coordinates of nodes on a graph. A wireless remote controller built using Android is an application that provides a 3d velocity. The updated coordinates get calculated using these velocity vectors, and the graph updates with the new coordinates of all nodes. The flow of control of a UAV's mobility in real-time is shown in Figure 5.

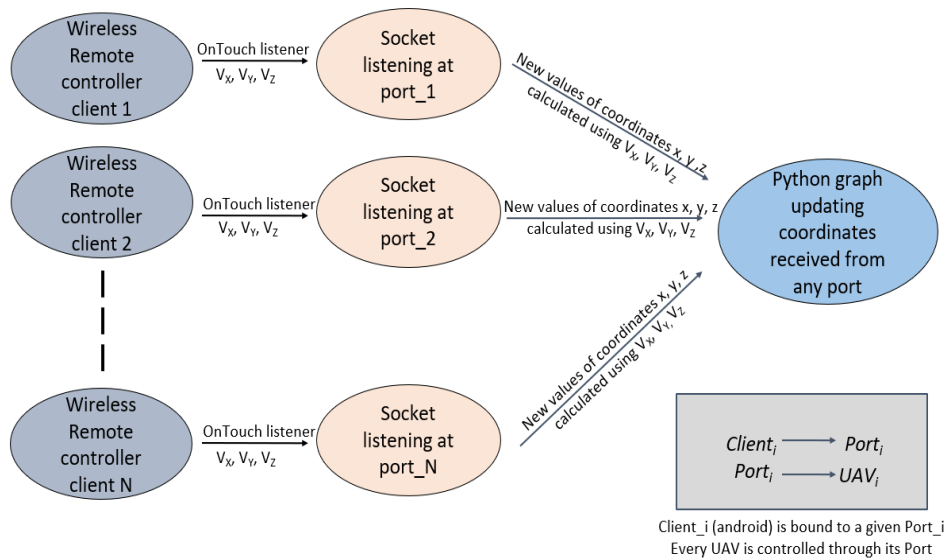

**Figure 5.** Flow of control describing the mobility of UAVs

---

**Algorithm 1** Registration of a UAV node

**Input**: UAV Public key
**Output:** New node registered upon successful registration
1: Registration Map: IoT ← UAV id
2: Blacklisted Map: Fault number ← UAV blockchain address
3: PubToMac Map: UAV id ← UAV public key
4: /*If registration function is triggered*/
5: **if** pubToMac[msg.sender]=0 **and** blacklisted[msg.sender]=0 **and** msg.value=5 ether **then**
6:     A new node is registered and initialised;
7: **else**
8:     **return**
9: /* function terminates */

10: /* If function remove faulty node is triggered */
11: **if** pubToMac[msg.sender]!=0 **and** registered[msg.sender].faulty!=0 **then**
12:     **if** registered[msg.sender]].penaltytoken=msg.value **and**
        (current_timestamp-registered[msg.sender].timestamp) <=
         registered[pubToMac[msg.sender]].faultytime) **then**
13:         The node is no more a faulty node
14:     **else if** msg.sender.timestamp > msg.sender.faulty
15:         $penaltytoken\ +\ 2$ ether **and** $faulty\ +\ 10$ seconds
16:     **else**
17:         **return**
18: /* function terminates */

---

**Algorithm 2** Transmission of data among nodes

**Input**: Destination address, data to be transmitted from the source
**Output**: If successful transmission: Data is received at the destination
         If unsuccessful transmission: Malicious node is found
1:     Mapping routeTable : node_address ← array of node address;
2:     /* if doTrans() function is triggered */
3:     **if** transaction=false **and** msg.sender!=dest **then**
4:         transaction ← true;
5:         source ← msg.sender

```
6:              destination ← dest;
7:              timestamp ← now;
8:         end if
9:    /*function terminates */

10:   /* Function RegisterCoordinates( ) begins when transaction is true */
11:   if transaction=true and node!=faulty then
12:      if node is registered and node!=GCS and paying registration amount then
13:             registered[pubToMac[msg.sender]].participating ← 1;
14:      else
15:             exit function;
16:      end if
17:      Add coordinates of the node to the registration hashmap
18:   end if
19:   /*function terminates*/

20:   /* function getTable( ) is triggered by user to investigate faulty and blacklist status of its node */

21:   /*If function pathFind is triggered*/
22:     BFS function is called
23:   if a path exists then
24:       return the route with minimum hops
25:   else
26:       return message "No route found"
27:   /*function terminates */
28:   /* function success( ) is triggered by destination upon successful transaction */
29:   if msg.sender=destination then
30:           Successful ← true;
31:   else
32:           return false;
33:   end if
34:   /* function terminates */
35:   /* function sendBackToken() triggered by successful transaction by destination */
36:   for i=0 to list.length-1
37:       if node participated in successful transaction then
38:            registered[list[i]].publicKey.transfer(1 ether);
39:       end if
40:   end for
41:   /* function terminates */
42:   /* function unsuccessful( ) is triggered upon unsuccessful transaction by destination */
43:   if msg.sender = destination then
44:       if data packet did not reach destination node then
45:            Faulty_node.participating ← 0;
46:            blacklisted[Faulty_node.id].publicKey]++;
47:            $Waittime of Faulty_{node} * 10$;
48:            $Penalty token of Faulty_{node} * 2$;
49:            Added as a culprit;
50:       else
51:         /*check data dissemination */
52:            for i=0 to i<route.length-1
```

```
53:            if current_node.data!=source_node.data then
54:             /* faulty node being the previous node not current */
55:               Faulty_node.participating ← 0;
56:               blacklisted[Faulty_node.id].publicKey]++;
57:               $WaittimeofFaulty_{node} * 10$;
58:               $PenaltytokenofFaulty_{node} * 2$;
59:               Added as a culprit;
60:               break;
61:            end if
62:          end for
63:     end if
64:     /* function terminates */
65:     /* function sendBackToken() return deposited token to intermediary nodes */
66:     /* function returnCulprit() returns address of culprit node */

67:     /* function transCompleted() is triggered by source upon successful transaction */
68:     if msg.sender =source and transaction=true and successful=true then
69:        for i=0 to route.length-2
70:             registered[Intermediary_node.id].publicKey.transfer(appreciation_token);
71:        end for
72:     end if
73:     /* function terminates */

74:      /* function send(data) triggered by node to send data */
75:     if msg.sender ∈ route nodes then
76:       if count=0 then
77:            Route[count].data ← string(x);
78:            Route[count].timestamp ← now;
79:       end if
80:       if count+1<=Route.length-1 then
81:            Route[count+1].data ← string(x);
82:             count++;
83:       end if
84:     end if
85:     /* function terminates */
86:     /* function getData() triggered by a node acquires data on a node of the route */
87:     /* function abort() can be triggered by only GCS */
88:     for i=0 to list.length-1
89:       if node=UAV and node.participating=1 then
90:          node.participating ← 0;
91:       end if
92:       routeTable[node].length ← 0;
93:       if node blacklist count <10 then
94:          node added to list1
95:       else
96:          Remove node from registration hashmap
97:       end if
98:       delete node from pubToMachashmap;
99:     end for
100:    delete list;
```

```
101:    for i=0 to count1
102:         Add nodes of list1 to registration hashmap
103:    end for
104:    /* function terminates */
```

# 5. Implementation Results and Details

This section will primarily discuss implementation details about the proposed simulation system.

## 5.1  Simulation of UAV network

To depict the simulation of a UAV network, we used matplotlib.pyplot [24] library. A 3-dimensional graph shows various UAV and GCS devices with their initial coordinates in this simulator. This graph is updated every five milliseconds to continuously update the coordinates and the condition of nodes in the network. The UAV network simulation is illustrated in Figure 6, 7, 8, and 9.

As shown in Figure 10, the wireless remote controller fetches the velocities of its UAV device and updates the coordinates accordingly. The UAVs are mobile devices, and they operate using controllers. All the UAVs have velocity vectors as Vx, Vy, Vz that are initially taken to 0, 0, 0, and based on the change in their velocities within a time period, the new coordinates are computed. The wireless remote controller uses socket programming for this purpose. Socket programming is used to send the velocity vectors from the client to the server in real-time. Every IoT device corresponds to a different port on the same server; for instance, UAV1 corresponds to port 8000, UAV2 corresponds to port 8001, and so on. Table 1 shows the color-coding scheme used in the simulation. As shown in Figure 11, a web page depicts the information related to various nodes (GCSs/UAVs) participating in the network.

## 5.2 Blockchain based network transactions

Blockchain implementation makes our system resilient to various cyber-attacks on a UAV network. Truffle [25] and Ganache [26] are used for implementing and deploying smart contracts in a blockchain network using Solidity (version >=0.4.21 <0.6.0) [27] language. Truffle is a development framework for Ethereum [28] that enables the user to develop, test and deploy smart contracts. It is an all-in-one platform for Solidity contracts that can deploy many public and private networks. Truffle provides the functionality of scriptable deployment, migration platform, and an interactive console for direct contract communication [25].

**Table 1:** Colour coding of simulation

| Network Actors / Condition | Colour Scheme |
|---|---|
| Ground Control Station | Blue Node |
| UAV | Black Node |
| Faulty UAV Node | Red Node |
| Transaction Successful | Green Node |
| Data Forwarding | Blue Dotted Lines |
| Dropped Data Packets | Black Dotted Lines |

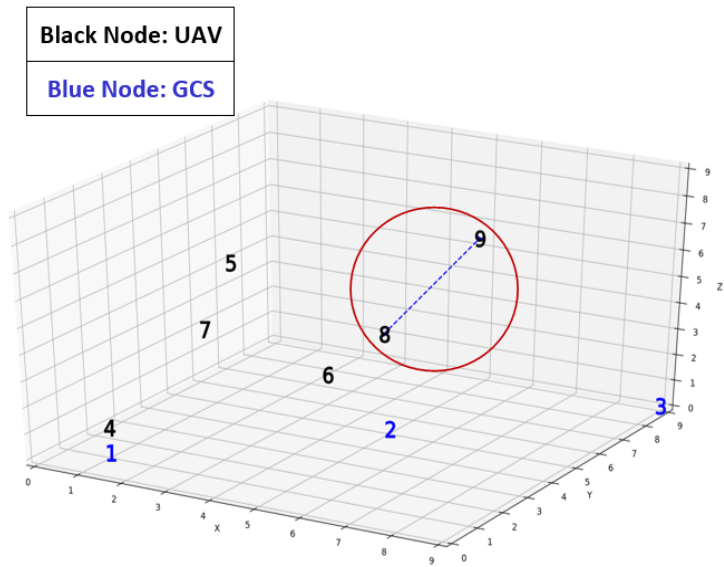

**Figure 6.** Data forwarding depicted using blue dashed line

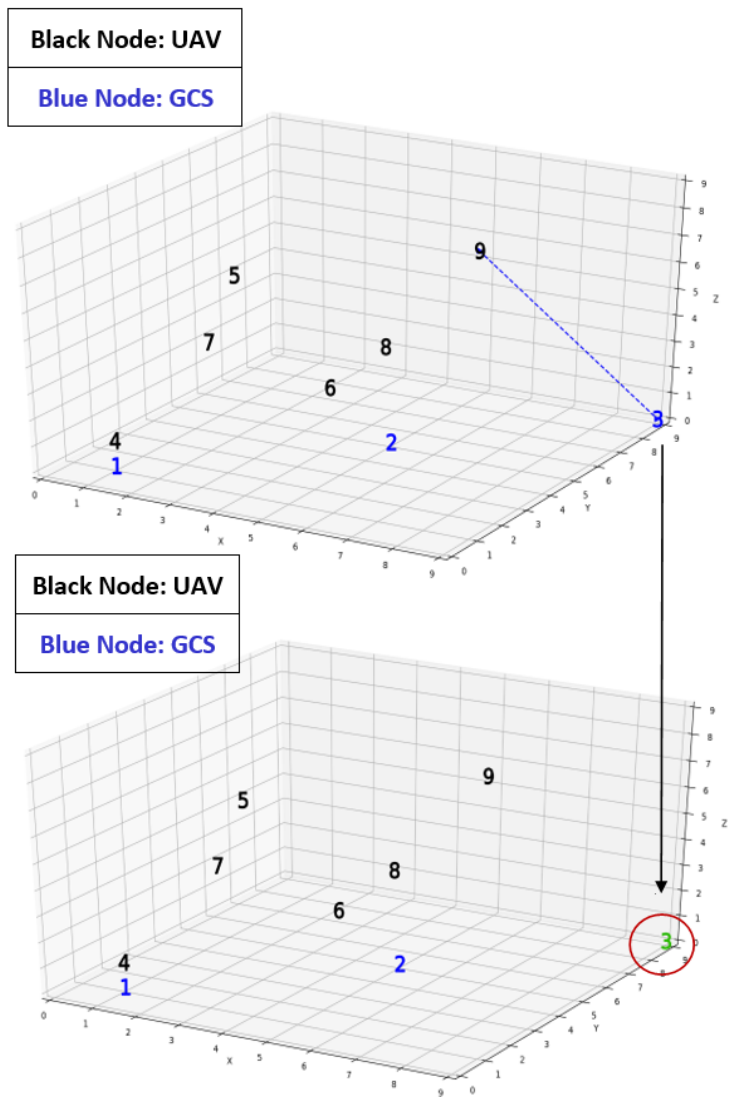

**Figure 7.** Destination (node 3) showing success of transmission by changing its colour to green.

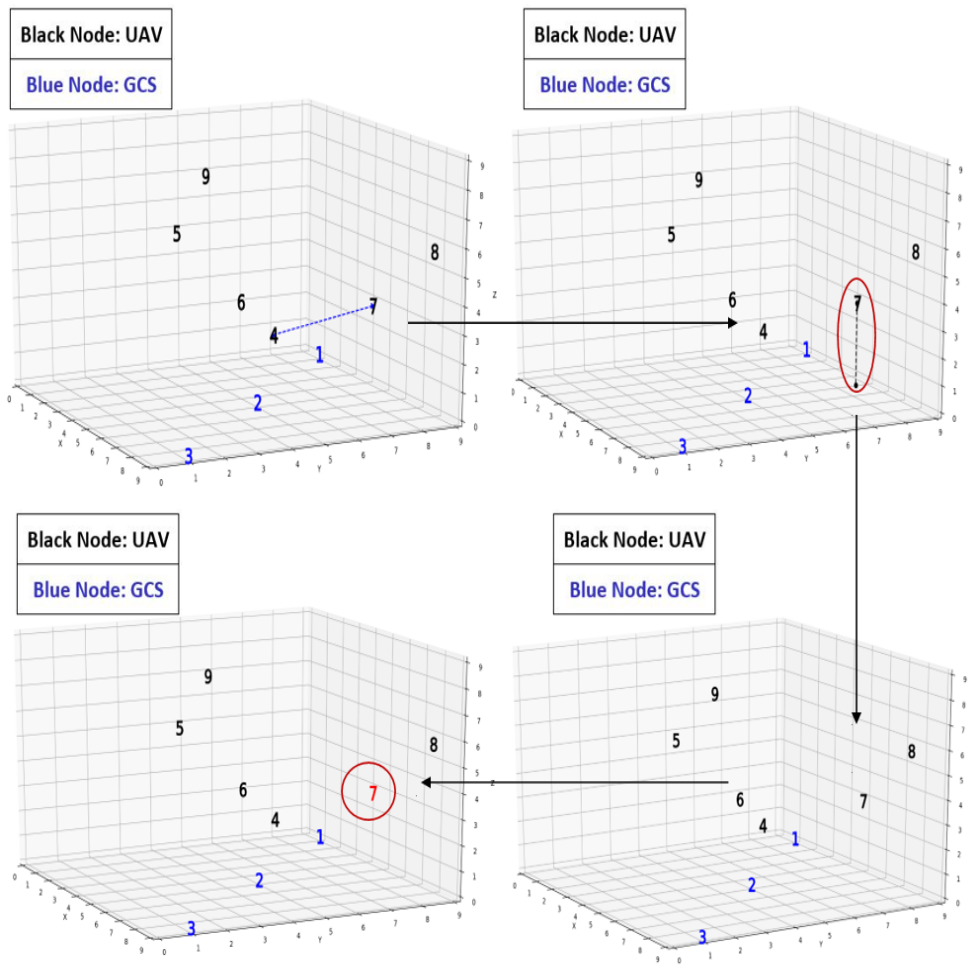

**Figure 8.** After dropping the packets (black dotted line), the malicious node becomes red after detection. The node again joins the network after paying the penalty token.

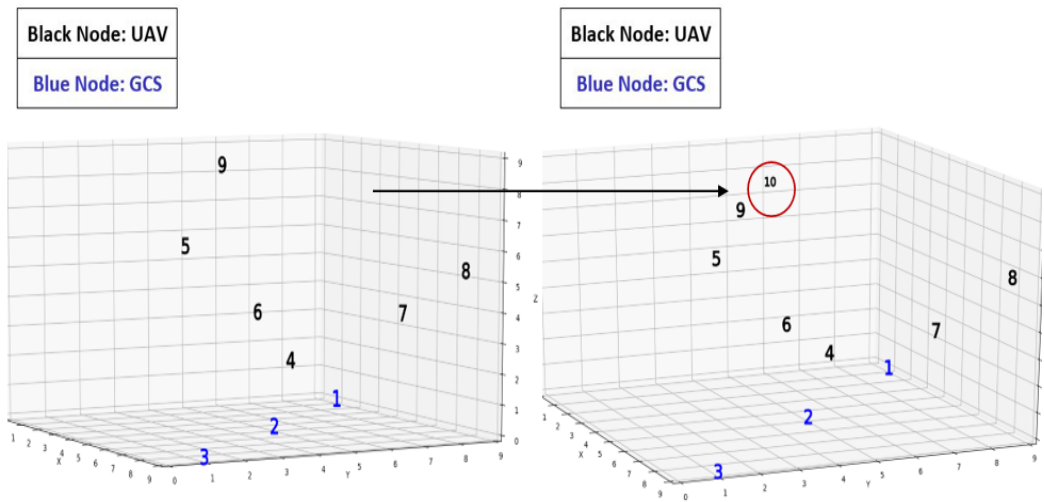

**Figure 9.** Registration of a new node

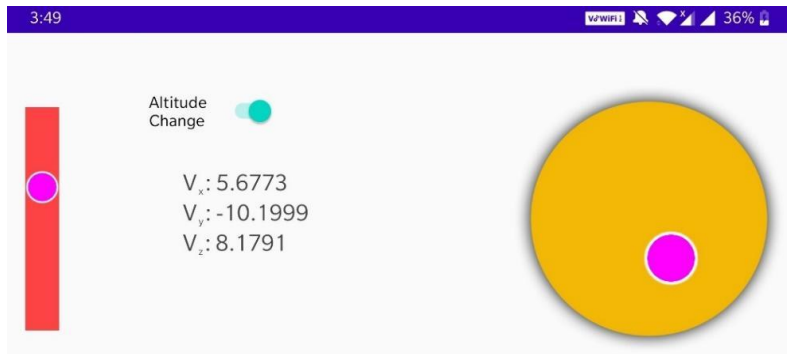

**Figure 10.** Android Wireless Remote controller with slider bar and altitude switch to provide velocities in the x, y and z direction.

| Contract Address: 0x30c4e0E00A8391AF8dC64149A77FDfc3eA1Ca42D | | | | |
|---|---|---|---|---|
| ADDRESS | BLACKLIST COUNT | FAULTY TIME | PENALTY TOKEN | BALANCE |
| 0x70E3638080c5F2d79d9ecDa07b915bCDA59efF50 | 0 | 0 s | 0 eth | 65.20 eth |
| 0x89d4046f422b59bb5CAB387d07d10126c7506F95 | 1 | 40 s | 2 eth | 100.00 eth |
| 0x921D10D6B1764f4bA35958a580086E1EC8D5Cae4 | 1 | 0 s | 10 eth | 99.82 eth |
| 0xC335ef44ac81d4D32c4b9636104865a21D4cDd77 | 10 | 20 s | 0 eth | 76.78 eth |
| 0x6DFBc4D1639e7707eACd3787066Ea8836f9E0B15 | 0 | 0 s | 0 eth | 84.93 eth |
| 0xB4E174b126Bc49f6b81341B54C08037D76d052E2 | 1 | 0 s | 0 eth | 84.95 eth |

**Figure 11.** A webpage showing real time information about the registered nodes

Ganache is a local blockchain development used when the user wants to develop a decentralized application on the Ethereum blockchain. It was previously called Testrpc, and it acts as a private blockchain that sets up ten default Ethereum addresses complete with private keys and pre-loads them with 100 simulated Ether each, as shown in Figure 12. The ganache is used to execute code on the simulated blockchain and, in turn, deploy smart contracts. Further, we use Web3 [29], a library that uses remote procedure call (RPC) communication to communicate with an Ethereum node. For interacting with the contracts deployed over the blockchain, this library is used to develop the user interface. We have created a decentralized platform for communication between the different network nodes.

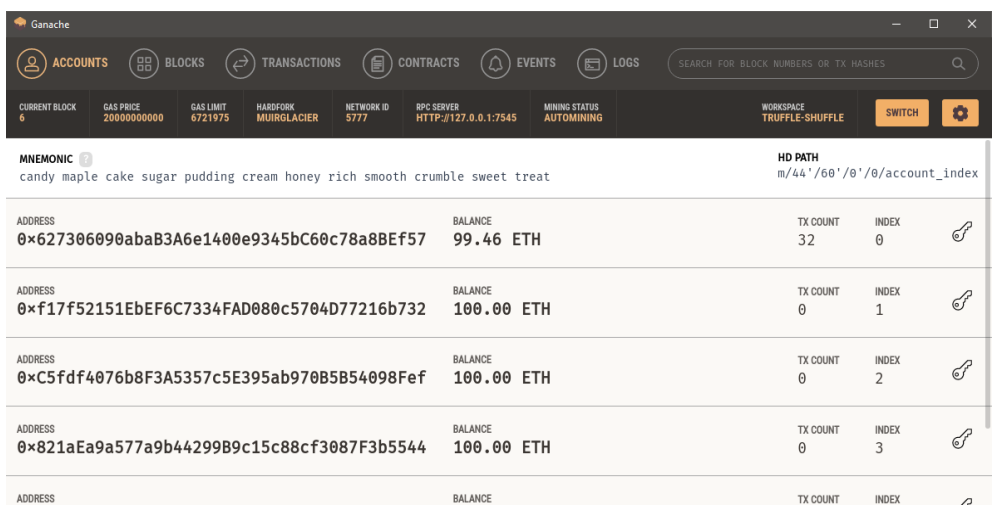

**Figure 12.** Several addresses with 100 Ethers each in Ganache.

Figure 13 illustrates Algorithm 1 that depicts the registration of a new node in the UAV network. When a UAV wants to register in the network, it is not already registered, not blacklisted. It pays the registration token; then, the UAV gets registered in the network. While registering the UAV node, the following parameters are taken & mapped in a registration hash map. Blockchain address of the UAV

- The public key of the UAV (used for data encryption)
- Fault time (time in which the suspected UAV requires to pay penalty tokens)
- Penalty token (extra amount of Ethers paid by a blacklisted node)
- Participating (bool value that indicates the participation of current node in the active transaction)
- Time Stamp (current time)
- x, y, z (coordinates of the UAV node)
- GCS (bool value that indicates whether the node is GCS or a UAV)

A faulty node trying to register in the network must first pay the penalty token within a given time frame provided. Failure to do so increases penalty tokens and the suspension of the faulty node from the network. A node is removed from the UAV network if blacklisted more than ten times.

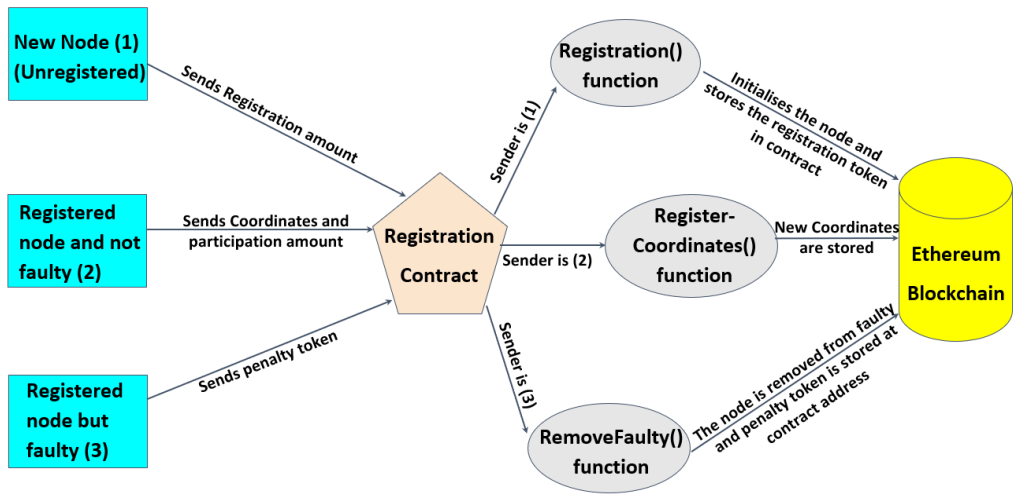

**Figure 13.** Registration of a UAV node

Figure 14 illustrates Algorithm 2 that depicts the transmission of data between nodes. When a node wants to initiate data transmission, it calls the doTrans() function of the DataSending contract. The function checks for any active transaction or if the destination address is the same as the senders. If both these conditions are false, the transaction initiates. Next, the nodes willing to participate in the network must register their current coordinates by calling the registerCoordinates() function. Only a registered UAV node is allowed to participate in the network and the coordinates of the unregistered nodes are not accepted. Also, it checks whether the node trying to register its coordinates is fault-free, and if found faulty, the node gets rejected.

Once all the nodes willing to participate in the transaction have registered, then the updateGraph() function is triggered, which updates the routing table that stores the neighbors of each node in the network. Each node checks whether all its neighbors are in the predefined distance range. The routing table is thus updated, and further, the pathFind() function gets called to find an optimal route for data transmission. This function uses the BFS algorithm to find the shortest route between the source and the destination.

Once an optimal path gets returned, the source node triggers the sendData() function for the next node in the given route. The data is encrypted using the public key and can be decrypted only by the private key of the destination. Hence, only the destination can access the data. Similarly, all the nodes transmit data further. Once the destination receives the data, it calls the success() function, which calls the sendBackToken() function, which returns the tokens contributed by the intermediary nodes to guarantee a trustful transaction. The source then calls the transCompleted() function that sends Ethers to the intermediary nodes as a token of appreciation for a successful transaction.

In case of a malicious node, the destination triggers the unsuccessful() function, which returns the submitted tokens to the intermediary nodes except for the culprit node. If the destination finds that the data received is corrupt, one can then discover the malicious node by comparing the data forwarded by each node with the data forwarded by the source node.

The abort() function can be triggered only by the base stations. This function aborts any active transaction in the network and removes blacklisted nodes from the network that are involved in malicious activities more than ten times.

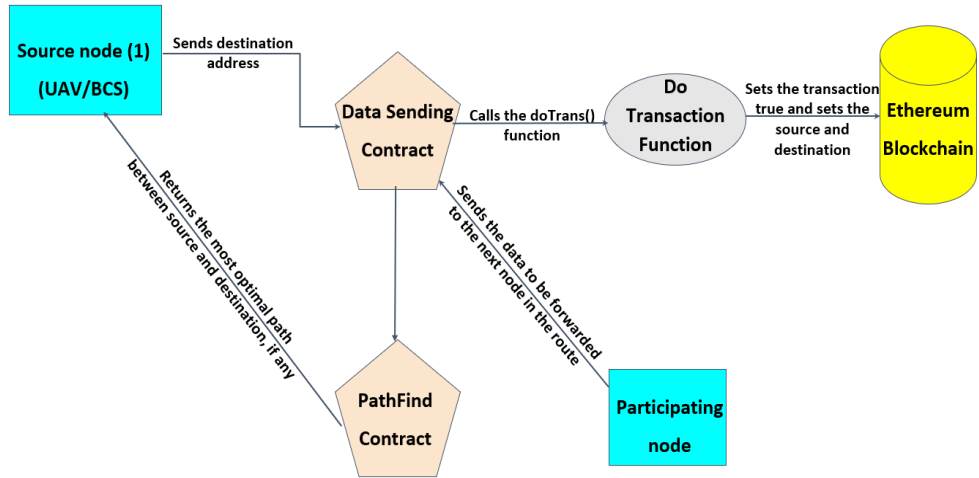

**Figure 14.** Transmission of data among nodes

## 6. Experimental Results and Discussions

We used the virtual blockchain provided by Ganache and compiled three contracts, namely, UAV.sol, DataSending.sol, and PathFind.sol, using the Truffle framework. For the simulation of the UAV network, we built a wireless remote controller to update the coordinates of the devices and their condition in the UAV network. We used the 'matplotlib.animation' to view the navigation and transmission of data. Web3 library made it possible to link the contracts deployed with the simulated system and interact with it. The virtual addresses provided by Ganache were pre-loaded with private and public keys. One hundred Ethers associated with each account for transactions in the Ethereum blockchain were used as the blockchain address for the UAVs.

For the simulation, an infrastructure with Ubuntu 20.04 OS with primary memory 8GB, secondary memory of 1 TB, and processing speed of 1.60 GHz is used. The designed system was tested for transmission of data over routes involving different numbers of intermediary nodes. The average time taken in the detection of attack from the time when the malicious node drops the packet is given by equation 5.

$$\beta = (n - 1 - x) * 2.9 \qquad (5)$$

where,
- $\beta$ = average time in seconds
- n = total number of nodes in the route
- x = number of intermediary nodes passed before dropping.

The results are obtained using the data shown in Table 2.

## 6.1 Detection and further Prevention of attacks

1. Blackhole attack: The route discovery is based on the BFS algorithm in our system. Therefore, selecting a route based on the advertised route by the malicious node is not possible. Also, once a node gets registered in the network, all the functions occur according to the deployed smart contracts. Since these contracts are immutable, so will detect any malicious activity by any node, and the node will be penalized.
2. Wormhole attack: In our system, a malicious node gets detected as soon as it performs a malicious activity; hence two attackers can't perform a colluding attack.

**Table 2:** Average Transmission delay among nodes

| No. of nodes in the route (n) | No. of nodes passed before dropping (x) | Time taken in detection of attack from the time when the malicious node drops the packet (sec) | Time for data transmission between 2 consecutive nodes (sec) |
|---|---|---|---|
| 8 | 3 | 11.6 | 3 |
| 9 | 4 | 11.6 | 3.1 |
| 10 | 4 | 14.5 | 2.8 |
| 7 | 4 | 5.8 | 2.9 |
| 6 | 4 | 2.9 | 3.1 |
|  |  |  | **Average time = 2.98 sec** |

3. Integrity attack: The prevention of data modification on reaching the destination gets decrypted using the destination's private key, which is unique for every node. The decrypted data gets compared with the data stored at a cloud database—the source node( before initiating the transaction, stores the data in this cloud database in an encrypted form). The malicious node is detected and penalized if data modification is found.
4. In DoS attack, the attacker selects a target node and makes it incapable of providing services. Here, every transaction occurs according to the logic defined in our smart contracts. Therefore, even if an attacker wants to flood a target node, it becomes impossible because any node in the route between the source and destination can only forward the data once. A node can only deliver a new data packet by initiating a new transaction in the blockchain. The decision to participate in the route solely depends upon the node, and hence, the attacker can't attack the target node by forcing the node to participate in the network.
5. Eavesdropping is a confidentiality attack where the node that is not the destination node accesses the confidential data. This attack can be prevented through the proposed system as the use of asymmetric cryptography ensures that no node other than the destination node can decrypt the data.

## 7. Conclusion and Future work

The potential usage of UAVs continues to increase day by day. In the future, smart cities will have UAVs playing a significant role in their development and functioning [40-42]. It can lead to the enhancement of services by businesses and franchisors. However, many have even started to adopt this technology after recognizing the incredible things that drones can do. UAV networks are prone to several attacks because these networks carry vital information. Their deployment requires private and reliable UAV communications; hence, it is essential to make the UAV network secure and resilient to cyber-attacks. This paper proposed a blockchain-based approach to make the UAV adhoc network secure and reliable. The use of blockchain provides data security and safeguards the network from the intrusion of malicious nodes. It also allows the GCS nodes and the UAV nodes to identify if tampering of data occurs. By creating a Python simulation of the UAV network with its functioning based on the smart contracts deployed in the Ethereum network, we could prevent several UAV network attacks such as blackhole attacks, gray hole attacks, DoS attacks, and data interception. We can also detect false information dissemination, and malicious node gets penalized. With the system proposed in this paper, we can improve the security and privacy of unmanned aerial vehicles in UAV-based adhoc networks. Possible future extensions of the designed system can be to provide different response systems for each type of attack in the UAV network [43-44].

**Biographical Statement**

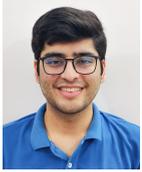
**Hardik Sachdeva** is a software engineer at Amazon, Dublin. He has 2+ years of working experience in the software engineering industry in various firms like Amazon, Goldman Sachs, Eka Care etc. He did his bachelors in Computer Engineering from National Institute of Technology, Kurukshetra in 2020. He has interest in research areas including Aerial networks, Blockchain, privacy and security of networks, UAV network, ad hoc and sensor networks, cyber security, cloud computing, internet of things.

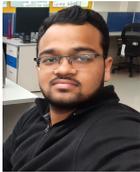
**Shivam Gupta** is a doctoral candidate in Computer Science at the Indian Institute of Technology (IIT), Ropar, with the Prime Minister Doctoral Fellowship from the Ministry of Education, Govt. of India. Interested in working on various topics, including Fairness in Machine Learning, Recommender Systems, Federated learning, Game Theory and Information Security. I finished my Bachelor's from the Indian Institute of Information Technology, Sonepat (Mentor & Campus NIT Kurukshetra). My research is published in premier computer science journals such as Springer, Taylor & Francis, etc. (https://www.web2geeks.in)

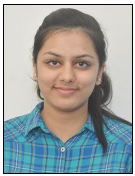
**Anushka Misra** is a software development Engineer at Amazon. She has 2+ years of working experience in various firms. She did her bachelors in Computer Engineering from National Institute of Technology Kurukshetra in 2020. She is interested in areas including cyber security, UAV network, Aerial networks, Blockchain, ad hoc and sensor networks, cyber security, cloud computing, internet of things.

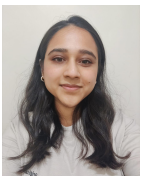
**Khushbu** is a software development Engineer at Expedia Group. She has 2+ years of working experience in various firms. She did her bachelors in Computer Engineering from Indian Institute of Information Technology Sonepat (Mentor and Campus NIT Kurukshetra) in 2020. She has interest in areas including Blockchain, network security, UAV network, Android applications, UX designing.

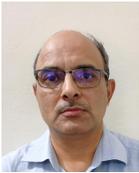
**Mayank Dave**

Prof. Mayank Dave received his Ph.D. in Computer Science and Technology from Indian Institute of Technology, Roorkee, India in 2002. He received B.Tech. degree from Aligarh Muslim University, Aligarh, India in 1989 and M.Tech. degree in Computer Science and Technology from IIT Roorkee, India in 1991. Member of Institutions of Engineers (India), Computer Society of India and IEEE USA, Dr. Dave is Professor in the Department of Computer Engineering at National Institute of Technology (NIT), Kurukshetra, India since 2013. He has so far guided 16 PhDs and has published over 200 research papers in various international/national journals and conferences. He has also guided 32 M.Tech. dissertations and over 100 B.Tech. projects. Prof. Dave has attended, presented papers and chaired technical sessions in several national and international conferences and seminars in India and abroad including USA, Italy, Singapore, China and Thailand. He has also served as Dean (Research and Consultancy) at NIT Kurukshetra during 2013-2016. He was Head of Department of Computer Engineering at NIT Kurukshetra for three terms and also Head of Department of Computer Applications of the Institute. Prof. Dave teaches Data Structures, Algorithms, Operating Systems, Java Programming and Computer Networks. His research interests include mobile ad hoc and sensor networks, cyber security, cloud computing, software defined networks, internet of things, cloud security etc. Prof. Dave also serves as a regular reviewer for many prestigious journals and conferences.